\documentclass[a4paper,11pt,headings=big,DIV=12]{scrartcl}
\usepackage{amsmath,amssymb}
\usepackage{color,xcolor}
\definecolor{blue}{rgb}{0,0,0.5} 
\usepackage[colorlinks,linkcolor=blue,urlcolor=blue,citecolor=blue]{hyperref}
\usepackage{graphicx}
\addtokomafont{disposition}{\rmfamily\boldmath} 
\usepackage[utf8]{inputenc}
\usepackage{enumerate} 
\usepackage{slashed}
\usepackage{cite} 
\usepackage{comment}
\usepackage{multirow}
 \usepackage{bbold}
\allowdisplaybreaks
\newcommand{\ba}{\be_a}
\newcommand{\bb}{\be_b}
\newcommand{\bc}{\be_c}
\newcommand{\barb}{\bar{b}}

\newcommand{\kd}{k_{\eps}}
\newcommand{\ONE}{\mathbb{1}}

\newcommand{\zam}[2]{\chi_{#1 #2}}
\newcommand{\zamS}[2]{\chi^{\MOM}_{#1 #2}}

\newcommand{\zamR}[3]{\chi_{#1 #2}^{#3}}

\newcommand{\Ra}{  {\RR_1} }
\newcommand{\RR}{  { {\cal R}} }
\newcommand{\Rb}{  { \RR_2} }

\newcommand{\RT}{  { \RR_{\chi}} }
\newcommand{\chischeme}{ \RR_{3\chi}}

\newcommand{\fin}{[\text{finite}]}

\newcommand{\MS}{{ \textrm MS}}

\newcommand{\als}{{a_s}}

\newcommand{\alsUV}{{a_s^\UV}}
\newcommand{\alsIR}{{a_s^\IR}}

\newcommand{\MOM}{{ \small \textrm{ MOM}}}

\newcommand{\lDSS}[1]{L_{#1 #1#1}^{\ONE,\MOM}}

\newcommand{\lDt}[1]{L_{#1 #1 #1}^{\ONE,\RR}}

\newcommand{\lnD}[2]{L_{#1 #2}^{\ONE,\RR}}

\newcommand{\lDtR}[2]{ L_{#1 #1 #1}^{\ONE,#2}}

\newcommand{\lnDR}[3]{ L_{#1 #2}^{\ONE,#3}}

\newcommand{\gm}{\mathrm{g}}

\newcommand{\TEMT}[1]{ T^{#1}_{ \;\;#1}}
\newcommand{\TEMTO}{\Theta}

\newcommand{\vev}[1]{\langle #1 \rangle}

\newcommand{\al}{\alpha}
\newcommand{\be}{\beta}
\newcommand{\ga}{\gamma}
\newcommand{\de}{\delta}
\newcommand{\la}{\lambda}
\newcommand{\eps}{\epsilon}

\newcommand{\Rtr}[2]{\Theta}

\newcommand{\UV}{{\small \textrm{UV}}}
\newcommand{\IR}{{\small \textrm{IR}}}

\newcommand{\Zpart}{{\cal Z}}

\newcommand{\SU}{\text{SU}}

\newcommand*{\mathcolor}{}
\def\mathcolor#1#{\mathcoloraux{#1}}
\newcommand*{\mathcoloraux}[3]{%
  \protect\leavevmode
  \begingroup
    \color#1{#2}#3%
  \endgroup
}

\begin{document}

\begin{flushright}
\begin{tabular}{l}
CP3-Origins-2018-027 DNRF90 \\
UUITP-30/18 \\
 \end{tabular}
\end{flushright}
\vskip1.5cm

\begin{center}
{\Large\bfseries \boldmath On the $a$-theorem in the Conformal Window}\\[0.8 cm]
{\Large Vladimir Prochazka$^a$
and Roman Zwicky$^b$,
\\[0.5 cm]
\small
$^a$  Department of Physics and Astronomy, Uppsala University,\\
Box 516, SE-75120, Uppsala, Sweden  \\[0.2cm]
$^b$ Higgs Centre for Theoretical Physics, School of Physics and Astronomy,\\
University of Edinburgh, Edinburgh EH9 3JZ, Scotland 
} \\[0.5 cm]
\small
E-Mail:
\texttt{\href{mailto:vladimir.prochazka@physics.uu.se}{vladimir.prochazka@physics.uu.se}},
\texttt{\href{mailto:roman.zwicky@ed.ac.uk}{roman.zwicky@ed.ac.uk}}.
\end{center}
  
\bigskip
\pagestyle{empty}

\begin{abstract}

We show that for four dimensional gauge  theories in the conformal window,
the Euler anomaly, known as the $a$-function, 
can be computed from a $2$-point function  of the trace of the energy momentum tensor  making 
it more amenable to lattice simulations. 
Concretely, we derive an expression 
for the $a$-function as an integral over the renormalisation scale of 
quantities related to $2$- and $3$-point functions of the trace of the energy momentum tensor. 
 The crucial ingredients are that the square of the field strength tensor is an exactly marginal operator 
 at  the Gaussian fixed point and that the relevant 
 $3$-point correlation function is finite when resummed to all orders.
This allows us to define a scheme  for 
which the $3$-point contribution vanishes, thereby explicitly establishing the strong version of the $a$-theorem 
for this class of theories. 
\end{abstract}

\newpage

\setcounter{tocdepth}{3}
\setcounter{page}{1}
\tableofcontents
\pagestyle{plain}

\section{Introduction}
\label{sec:intro}
 
The conformal anomaly, first known as the central charge  $c$ of the Virasoro algebra, 
is key to the  physics of conformal field theories (CFTs) as it is a measure 
of the number of degrees of freedom. 
The Weyl anomalies discovered in the 70's (see \cite{20years} for a review of the topic) state that this central charge 
appears in the trace of the energy momentum tensor (TEMT) when there 
is a curved background, $\vev{\TEMT{\rho}}_{\textrm{CFT}} = - (\be_c/(2 4 \pi)) R$, 
elevating the central charge to a $\be$-function; $\be_c \equiv c$.  
The $c$-theorem \cite{Zamolodchikov:1986gt} can be stated in terms of Cardy's formula \cite{Cardy:1988tj} 
\begin{equation}
\label{eq:mom2D}
\Delta \be_c^{2D} = \be_c^\UV -  \be_c^\IR = 3 \pi \int d^2 x \, x^2  \vev{ \TEMTO(x) \TEMTO(0)}_c 
\geq 0 \;,
\end{equation}
where $\TEMT{\rho}|_{\textrm{flat}} \to \TEMTO$ and $\vev{\dots}_c$ stands for the connected component of the vacuum expectation value (VEV). 
It is assumed that the theory flows from an ultraviolet (UV) to an infrared (IR) fixed point (FP).
The inequality  $\Delta \be_c^{2D} \geq 0$, establishes the irreversibility of the renormalisation 
group (RG) flow, and follows from the positivity of the spectral representation and the finiteness of the correlator in \eqref{eq:mom2D}.  

In 4D the situation is more involved as  there are further terms in the TEMT
 \begin{equation}
\label{eq:VEVTEMT}
\vev{\TEMT{\rho}(x)} 
 =   -( \ba^\IR E_4 + \bb^\IR H^2 + \bc^\IR W^2  )  + 4 \barb^\IR \Box  H 
  \;.
\end{equation}
Above 
 $H \equiv R/(d-1)$ and  as opposed to  \cite{PZboxR} we have omitted a cosmological constant term 
for brevity. 
In particular, we denote the coefficients of the geometric invariants by $\be$-functions 
except the $\Box R$-term which is a Weyl variation of the local $R^2$-term. 
In CFTs, $\be_b =0$ and $\be_a$, $\be_b$ \& $\barb $ are the true conformal anomalies. 

The $a$-theorem, $\Delta \ba =  \ba^\UV -  \ba^\IR  > 0$, was conjectured early on
\cite{Cardy:1988cwa} and a proof in flat space uses the  $4$-point function of the TEMT, 
anomaly matching and analyticity  \cite{KS11,K11}.\footnote{In curved space, $\be_a$ can be assessed 
from a $2$-point function \cite{FL98}.}
A stronger version of the theorem requires an interpolating function $\tilde{\be}_a(\mu)$ that reduces to $\ba^{\UV,\IR}$ at the respective FPs, satisfying  monoticity, $d_{\ln \mu} \tilde{\be}_a  \geq 0$, along 
the RG flow.  A  perturbative argument was given in 
 \cite{Osborn:1989td} by finding a function satisfying $d_{\ln \mu} \tilde{\be}_a  = (\chi_{AB} + \dots)\be^A \be^B$
where the Zamolodchikov metric $ \chi_{AB}$ is positive by unitarity at the Gaussian FP.\footnote{This argument was generalised to conformal perturbations at interacting FPs in \cite{BKZR14}. In both cases the positivity is controlled by the smallness of perturbative corrections encoded in the dots. In 2D the strong $c$-theorem was proven in the original paper \cite{Zamolodchikov:1986gt} without reference to perturbation theory.} 

\begin{figure}[t]
\centering
\includegraphics[scale=0.60]{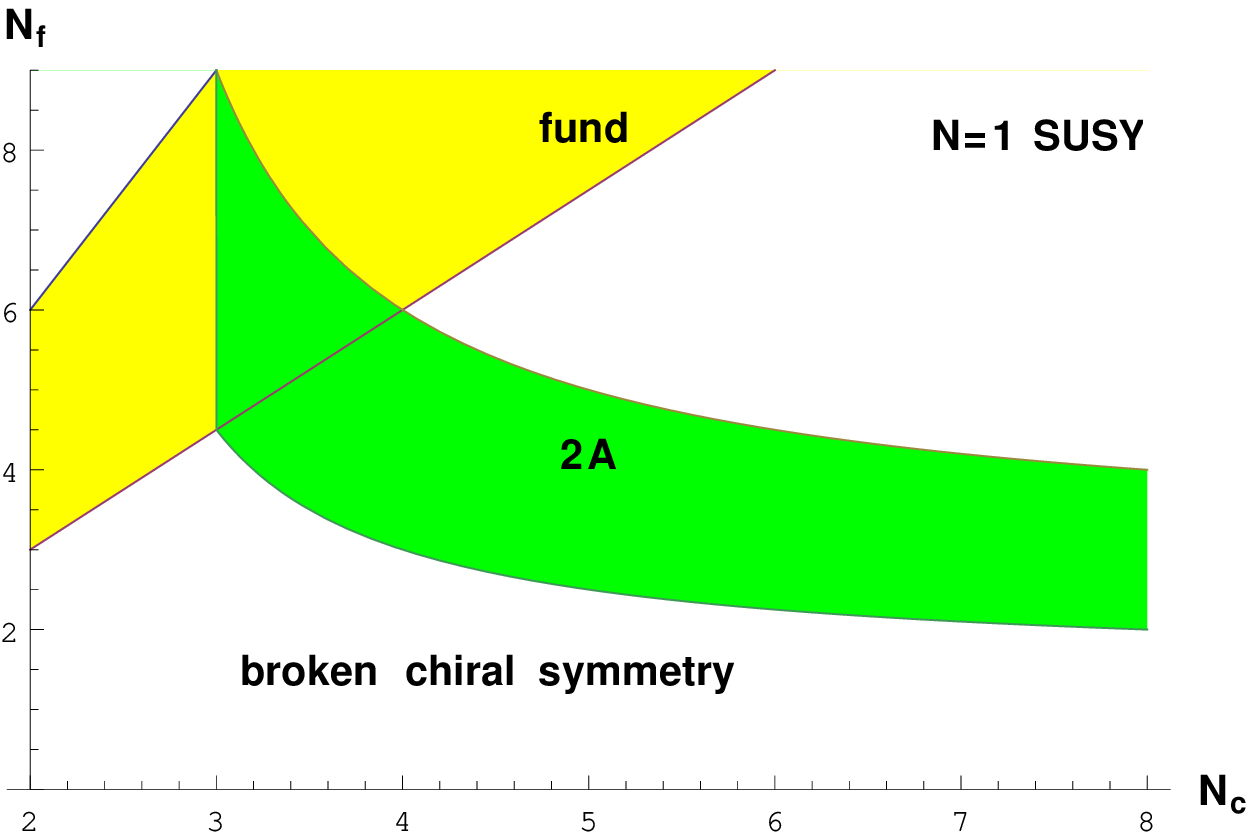}\includegraphics[scale=0.60]{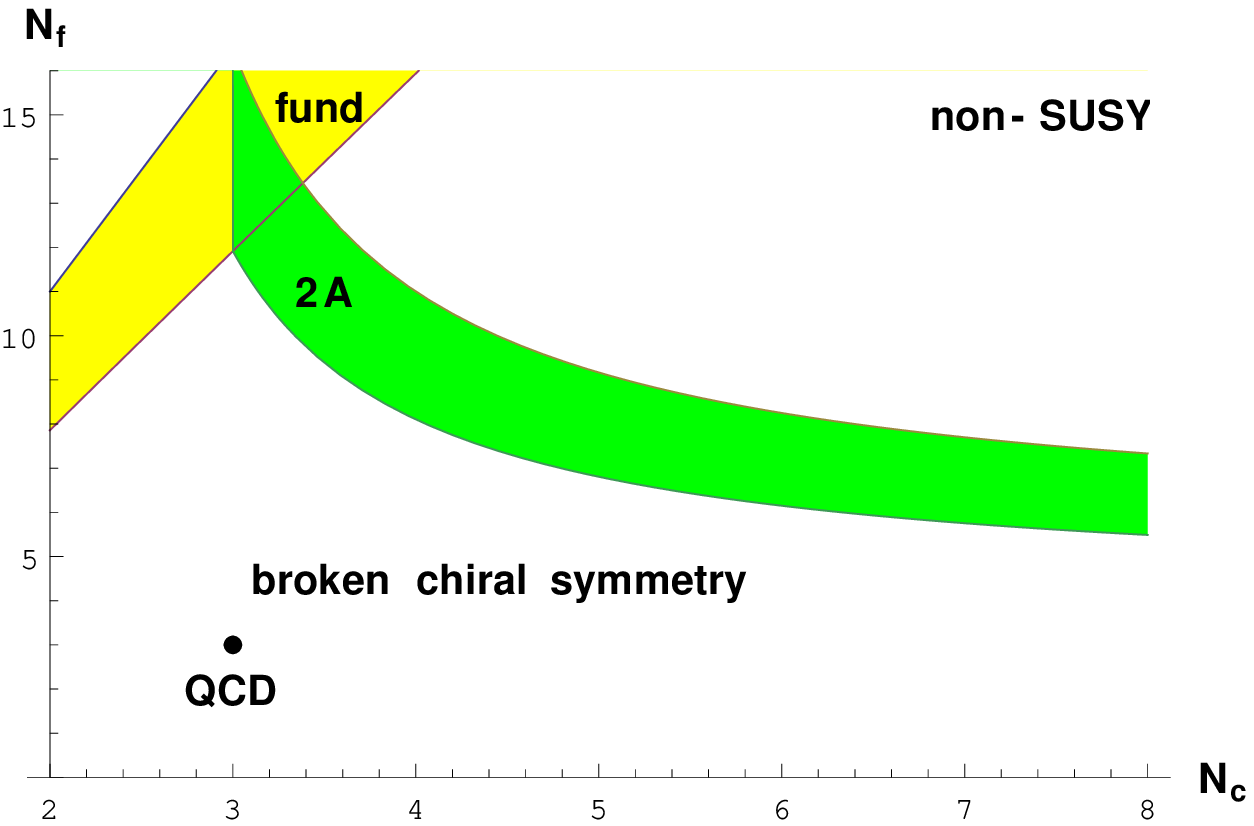}
\caption{\small The conformal window for supersymmetric (left) and non-supersymmetric  (right) gauge 
theories for quarks in the matter in the fundamental (yellow) and $2$ index antisymmetric  (green) representation of the $SU(N_c)$ gauge group.  
The upper boundaries are dictated by the loss of  asymptotic freedom and the lower boundaries 
are known in ${\cal N}=1$ supersymmetric gauge theories thanks to the electric-magnetic duality \cite{Intriligator:1995au} and 
for non-supersymmetric gauge theories they are debated in lattice simulations and for the actual values we 
have taken the boundaries given by Dyson-Schwinger equations \cite{Hill:2002ap}. 
Inside the yellow and green bands the theory is expected to flow to an conformal IR FP. Below these regions the 
chiral symmetry is spontaneously broken in the IR which is the case for Quantum Chromodynamics (QCD).}
\label{fig:CW}
\end{figure}

A representation similar to \eqref{eq:mom2D} has been proposed involving 
 $2$- and $3$-point functions  \cite{A03}  
\begin{eqnarray}
\label{eq:3ptSumRuleBa}
\Delta \ba &\;= \;&  \frac{1}{3 \cdot 2^8}\left(   \int_x x^4 \vev{\TEMTO(x) \TEMTO(0)}_c - 2
\int_{x} \int_y  \ [ (x \cdot y)^2 - x^2y^2] 
\vev{\TEMTO(x)\TEMTO(y)\TEMTO(0)}_c \right) \;,
\end{eqnarray}
where $\int_x \equiv \int d^4 x$. This expression is derived in appendix \ref{app:sumrules}
using conformal anomaly matching. As alluded above,
this expression does not lend itself to positivity because of the presence of the 
$3$-point  function. In this work, we will show that for gauge theories with gauge couplings 
only, the $3$-point function term drops for  theories in the conformal window cf. fig~\ref{fig:CW}. This establishes 
the positivity with Euclidean methods and makes the evaluation more amenable to 
lattice simulations. 

\subsection{Executive Summary}

In the remainder of this introduction, we give an executive summary of our work leaving 
the derivation of equations and definitions of schemes to the main part of the paper.
Our assumptions are: (i) that the TEMT assumes the form 
\begin{equation}
 \label{eq:TraceTen}
\TEMTO \sim \be^A [O_A]  + \textrm{equation of motion terms} 
\quad  (\Leftarrow \quad {\cal L} = g^A_0 O_A) \;,
\end{equation}
(summation over $A$ implied) and (ii) that the beta functions 
 $\be^A \equiv \frac{d}{d \ln \mu} g^A$ vanish in the IR \& UV.\footnote{For gauge theories with 
chiral symmetry breaking, the assumption $\TEMTO \sim \be^A O_A$ breaks down 
since the goldstone bosons couple with a term $\TEMTO \sim \Box \pi^2$ which cannot be 
improved since it is in conflict with chiral symmetry \cite{LS89,DL91} leading to subtleties for flow theorems \cite{LPR12,PZboxR}.} 
Above operators with square brackets denote renormalised 
composite operators, e.g. $O_A \sim G^2$ in the case at hand,
 where $G^2 \equiv (G_{\mu \nu}^a)^2$, is the standard field strength tensor squared known from quantum chromodynamics (QCD).
Using these assumptions allows us to derive
\begin{eqnarray}
\label{eq:DelBa1}
\Delta \ba &\;=\;&
  \frac{1}{4} \int_{-\infty}^{\infty}( \zamR{A}{B}{\RR}(\mu') \be^A \be^B -  \zamR{A}{BC}{\RR}(\mu') \be^A \be^B  \be^C)d\ln \mu'  \;,
\end{eqnarray} 
from \eqref{eq:3ptSumRuleBa}.
 The $\be$-functions are $\mu$-dependent  through the couplings and 
the  $\chi$'s are the analogues of the Zamolodchikov metric (cf. appendix \ref{app:Conventions} for
 definitions and notational conventions).\footnote{Whereas the $\chi$'s
 are  dependent on  a generic scheme $\RR$, the two flow integrals themselves are scheme 
 independent, cf.  \cite{PZboxR}. We will refer to $\chi_{ABC}$ as the $3$-metric throughout in a loose
 analogy to the Zamolodchikov metric in two dimensions.} The expression 
 \eqref{eq:3ptSumRuleBa} is derived in section \ref{sec:sumrules}, and is a new result of this paper.
 
 For our work the crucial input is that the leading order correction to the non-interacting FP is
 \begin{equation}
 \label{eq:chi23}
\zamR{A}{BC}{\RR} ={\cal O}((g^I)^2) \;, \quad     \zamR{A}{B}{\RR} ={\cal O}((g^I)^0)  \;.  
 \end{equation}
The main focus of this paper will be on asymptotically free QCD in the conformal window regime where \eqref{eq:chi23} follows by using the conformal OPE (c.f. appendix 
  \ref{app:formal} ) and is of course easily established by direct computation as well. 
 Using \eqref{eq:chi23} and our previous work on finiteness of $2$- and $3$-point functions we
 are  able to define a scheme, referred to as the $\chischeme$-scheme, for which 
 $ \zamR{A}{BC}{\chischeme}(\mu) =0$ along the flow. This establishes the main result of our paper
   \begin{eqnarray}
\label{eq:DelBaProved}
\Delta \ba &\;=\;& 
  \frac{1}{4} \int_{-\infty}^{\infty}( \zamR{A}{B}{\RR}(\mu') \be^A \be^B  )d\ln \mu'  =    \frac{1}{3 \cdot 2^8}   \int_x x^4 \vev{\TEMTO(x) \TEMTO(0)}_c  > 0 \;,
\end{eqnarray} 
valid for the assumptions specified above and satisfying \eqref{eq:chi23}. 
On a side note this means that $\Delta \ba = 2 \Delta \barb  \equiv 2( \barb^\UV - \barb^\IR)  $, for the same conditions, 
since the flow theorem 
for $\barb$ can be expressed in terms of the same $2$-point function \cite{PZboxR}. This relation was conjectured to hold for general classically conformal QFTs in \cite{A99}. In this work we show 
under what conditions this relation holds. A case where it fails is when  the theory contains scalar couplings and the 3-point function does contribute.

The paper is organised as follows.  
The cornerstones, formula \eqref{eq:DelBa1} and the scheme $\zamR{A}{B C}{\RT}(\mu) =0$, 
are established in sections   \ref{sec:sumrules} and  \ref{sec:three} respectively.  
More precisely, in section \ref{sec:vanishingUV} it is shown that $\zamR{A}{B C}{\RT}$ satisfies 
\eqref{eq:chi23}, used in section \ref{sec:finite3} to derive  the  finiteness of the counterterm which 
then allows the explicit construction of the scheme for which  $\zamR{A}{B C}{\RT}(\mu) =0$ 
in section \ref{sec:scheme}.
Definitions, including notation from our previous work, are reviewed in appendix \ref{app:Conventions}.
The sum rule \eqref{eq:3ptSumRuleBa} is derived in appendix \ref{app:sumrules} and a more
formal argument for the first equation in \eqref{eq:chi23}, underlying the above mentioned scheme, is given in appendix \ref{app:formal}.

\section{The Flow of $E_4$ (or $\ba$) as an Integral over the RG-scale}
\label{sec:sumrules}

It is the aim of this section to derive \eqref{eq:DelBa1}. 
 The presentation below is similar to the one given 
for the $\Box R$-flow in \cite{PZboxR} where we have shown that\footnote{For remarks
concerning adding a local term $\de {\cal L} \sim w_0 R^2$ to the bare action, with regards 
to eqns \eqref{eq:3ptSumRuleBa}  and \eqref{eq:Dbarb}
 cf. 
appendix \ref{app:DilatSumRule} or our previous work \cite{PZboxR}.}
\begin{equation}
\label{eq:Dbarb}
\Delta \barb =   \frac{1}{3 \cdot 2^7}   \int_x x^4 \vev{\TEMTO(x) \TEMTO(0)}_c = 
   \frac{1}{8} \int_{-\infty}^{\infty}  \zamR{A}{B}{\RR} \be^A \be^B  d\ln \mu'  \;.
   \end{equation}
   The starting point is formula \eqref{eq:3ptSumRuleBa}, which is derived in appendix 
\ref{app:sumrules}. Using \eqref{eq:Dbarb} one may write 
   \begin{equation}
\label{eq:BaFlow1}
\Delta \ba = 2 \Delta \barb -\frac{1}{4}\hat{P}_{\la_3} \Gamma_{\theta \theta \theta}(p_x,p_y) |_{p_x=p_x=0} \;,
\end{equation}
  where 
\begin{equation}
\label{eq:FTttt}
\Gamma_{\theta \theta \theta}(p_x,p_y)  =   
 \int_{x} \int_y  e^{i (p_x \cdot x + p_y \cdot y ) } \vev{\TEMTO(x) \TEMTO(y) \TEMTO(0)}_c \;,
\end{equation}
and  $\hat{P}_{\la_3}$ is defined in \eqref{eq:3ptProjector}.\footnote{Assuming the IR limit $p_x,p_y \to 0$ to be regular, we can choose to approach $0$ by taking, for example, $p_x=-p_y=p \to 0$ and define a function 
$f(p^2)  \equiv \hat{P}_{\la_3} \Gamma_{\theta \theta \theta}(p,-p)$.}

The transformation of the equation above into an integral representation 
over the RG scale, necessitates the discussion of the renormalisation prescription. To regularise we use dimensional regularisation with $d=4-2\eps$. 
The correlator is renormalised by
a splitting of  the bare function into a renormalised $\Gamma^\RR$ and a counterterm $L^\RR = \sum_{n \geq 1} 
L_n^\RR \eps^{-n} $
\begin{eqnarray}
\label{eq:split}
\Gamma_{\theta \theta\theta }(p_x,p_y) =
 \Gamma_{\theta \theta\theta }^\RR(p_x,p_y,\mu)  + 
 \lnD{(\theta) }{\theta \theta}(\mu)  P_3 + 
 \lnD{\theta }{\theta \theta}(\mu) \lambda_3 \; ,
 \end{eqnarray}
 which consists in a Laurent series.
Above $\lambda_3$ (K\"all\'en function) and $ P_3$ are the $3$- and $2$-point kinematic structures 
\begin{alignat}{2}
\label{eq:kine}
& \lambda_3&\;\;\equiv\;& p_x^4+p_y^4+p_z^4-2(p_x^2 p_y^2 + p_x^2 p_z^2+ p_y^2 p_z^2)  \;, \nonumber \\[0.1cm]
& P_3  &\;\;\equiv\;&  p_x^4  + p_y^4+p_z^4 \;,
\end{alignat}
where momentum conservation, $p_z +  p_x+p_y = 0$,  is implied. The quantities $\lnD{(\theta) }{\theta \theta}(\mu)$,$ \lnD{\theta }{\theta \theta}(\mu)$ are Laurent series in $\eps$ depending on the running couplings of the theory. From \eqref{eq:BaFlow1} it is seen that  $ \lnD{\theta }{\theta \theta}$ is the key quantity which 
we analyse by its scale dependence 
\begin{equation}
\label{eq:ZamThetaDef}
 \zamR{\theta \theta}{\theta}{\RR}(\mu)  =  
\left(  2\eps -\frac{d}{d \ln \mu} \right) \,   \lnD{\theta }{\theta \theta}(\mu) \stackrel{\eps \to 0}{=}- \frac{d}{d \ln \mu} \lnD{\theta }{\theta \theta}(\mu) \;.
\end{equation}  
In the last equality we used the result of \cite{PZfinite} that $\lnD{\theta }{\theta \theta}$ is finite after resummation of divergences. \\
In  establishing the flow formula \eqref{eq:DelBa1}, we follow the logic of \cite{PZboxR} and introduce the so-called \MOM-scheme, defined by
\begin{equation}
\label{eq:Zamo3}
\zamS{\theta}{\theta \theta} = - \frac{d}{d \ln p} \Big|_{ p= \mu} \hat{P}_{\la_3} \Gamma_{\theta \theta \theta}(p,-p) \;.
\end{equation}
By solving the above ODE we arrive at 
 \begin{eqnarray}
\label{eq:Cscheme3}
\hat{P}_{\la_3} \Gamma_{\theta \theta \theta}(g^Q(p)) &\;=\;& \int^\infty_{\ln p/\mu_0}  \zamS{\theta}{\theta \theta}  d\ln \mu'
\nonumber \\[0.1cm]
&\;=\;&  \underbrace{\int^{\ln \mu/\mu_0} _{\ln p/\mu_0}  \zamS{\theta}{\theta \theta}  d\ln \mu'}_{\hat{P}_{\la_3} \Gamma^\MOM_{\theta \theta \theta}(p/\mu, g^Q(\mu))} + 
\underbrace{\int^\infty_{\ln \mu/\mu_0}  \zamS{\theta}{\theta \theta} d\ln \mu'}_{\lDSS{\theta}(g^Q(\mu))} \;,
\end{eqnarray}
where the split on the second line is compatible with \eqref{eq:ZamThetaDef}.
In order to pass to the coupling coordinates one uses the assumption \eqref{eq:TraceTen} to write 
\begin{equation}
\Gamma_{\theta \theta \theta}=\be^A\be^B\be^C\Gamma_{A B C}(p_x,p_y) \;,
\end{equation}
 with
\begin{equation} \label{eq:GammaABC}
\Gamma_{A B C}(p_x,p_y)  =   
\int_{x} \int_y  e^{i (p_x \cdot x + p_y \cdot y) } \vev{[O_A(x)][ O_B(y)][  O_C(0)]}_c \;.
 \end{equation}
Renormalisation of the above correlator and further  definitions (e.g. 
$\chi_{ABC}^\RR$ in \eqref{eq:ChiABC})
are reviewed in appendix \ref{app:Conventions}. It is now possible to define \MOM-scheme relation analogical to \eqref{eq:Zamo3} for the correlator \eqref{eq:GammaABC}. In this \MOM-scheme we can use the relation  $ \zamS{\theta}{\theta \theta} = \be^A \be^B \be^C \zamS{A}{B C}$, which in turn follows from substituting \eqref{eq:TraceTen} into \eqref{eq:Zamo3}.

The final expression for $\Delta \ba$ is obtained by taking $p\to 0$ limit of \eqref{eq:Cscheme3} and inserting the result to \eqref{eq:BaFlow1}
\begin{eqnarray}
\label{eq:DelBa2}
\Delta \ba &\;=\;&
  \frac{1}{4} \int_{-\infty}^{\infty}( \zamS{A}{B} \be^A \be^B -  \zamS{A}{B C} \be^A \be^B  \be^C)d\ln \mu'  \;.
\end{eqnarray}
Although as it stands \eqref{eq:DelBa2} is written in a specific scheme, just like for $\Delta \barb$ \cite{PZboxR},  
scheme-independence follow by observing that a change from a scheme $\Ra$ to $\Rb$ 
is given by a cohomologically trivial term
 \begin{equation}
 \label{eq:scheme-change3}
  \de \zam{\theta}{\theta\theta}   =  \zamR{\theta}{\theta\theta}{\Rb}  -  \zamR{\theta}{\theta\theta}{\Ra} =   \frac{d}{d \ln \mu} \omega \;,
 \end{equation} 
 where $\omega = (\be^A \be^B \be^C \omega_{ABC})$  with $\omega_{ABC}$ parametrising 
 the  change of scheme cf. eq.~\eqref{eq:Zam3scheme}. This establishes 
 the representation \eqref{eq:DelBa1} and completes the aim of this section. 
 We have also checked that eq. \eqref{eq:DelBa2} is consistent with the $\MS$-scheme formulae of  \cite{JO90} (namely 3.17b and 3.23 in this reference).
  
\section{The $3$-metric $\zam{g}{gg}$ in Gauge Theories}
\label{sec:three}
In this section we  restrict ourselves to QCD-like theories with one gauge coupling $g$ and massless fermions. The generalisation of the following result to multiple coupling theories satisfying \eqref{eq:chi23} is straight forward. Before we proceed, let us establish the notation. The trace anomaly for gauge-theories reads
\begin{eqnarray} \label{eq:TraceTenQCD}
\Theta= \frac{\be}{2} [O_g] \; ,
\end{eqnarray}
where $\be= \frac{d \ln g}{d \ln \mu}$ is the logarithmic beta function.
The corresponding operator $O_g$ is 
the field strength squared
\begin{equation}
\label{eq:Og}
[O_g] = [\frac{1}{g_0^2} G^2]  \;,
\end{equation}
with the somewhat non-standard treatment of the coupling constant (and $G^2$ has been defined previously). The mapping to the general expressions \eqref{eq:DelBa1} and \eqref{eq:DelBaProved} is done by comparing \eqref{eq:TraceTen} and \eqref{eq:TraceTenQCD}, which gives $O_A \to O_g$ and 
$\be^A \to \frac{1}{2} \be$ omitting the superscript $g$ on the $\be$-function for brevity.

In section \ref{sec:vanishing} it is shown that $\chi_{ggg} ={\cal O}(g^2)$, from 
where it is deduced, in section \ref{sec:finite3}, that the counterterm to the $3$-point function $L_{ggg}$
is finite when summed to all orders for \emph{all} points along the flow.  Based on this, in section \ref{sec:scheme}, 
 a scheme is defined for which $\chi_{ggg}^{\chischeme}(\mu) = 0$,  leading to the main result of the paper.

\subsection{The vanishing of $\chi_{ggg}$ at the UV Fixed Point} \label{sec:vanishing}
\label{sec:vanishingUV}

\begin{figure}[t]
\centering
\includegraphics[scale=0.60]{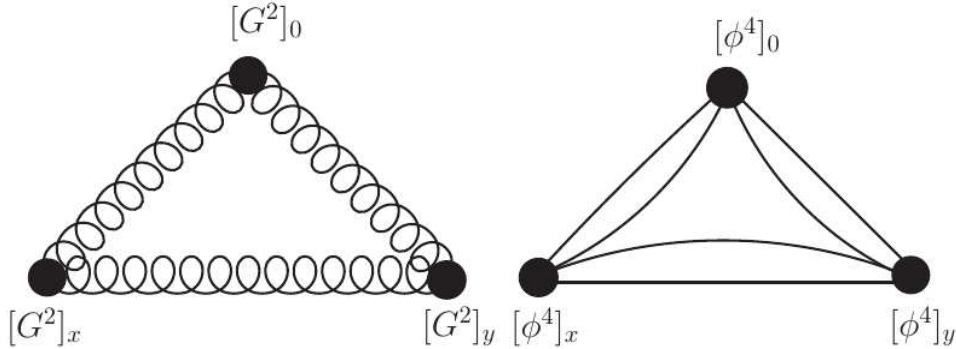}
\caption{\small (left) Leading order diagram contributing to $\Gamma_{ggg}$ with no divergence 
in the $3$-point structure $\la_3$ after Fourier transformation. This is in accordance with \eqref{eq:chi23}. (right) Leading order diagram contributing to the correlator of $\phi^4$-operators \eqref{eq:phiphiphi}. In momentum space this corresponds to a four loop graph and does lead to divergencies in the $\la_3$-structure 
from where one can infer that the $\phi^4$ coupling acquires a RG running.}
\label{fig:gluon}
\end{figure}

Technically, we will show that $\chi_{ggg} ={\cal O}(g^2)$. 
The $3$-point function can be computed at leading order directly in momentum space by evaluating the diagram in fig.~\ref{fig:gluon} which gives
\begin{eqnarray}
\label{eq:GGG}
\Gamma_{ggg}(p_x,p_y)  &\;\;\equiv\;&  \frac{1}{g_0^6}
 \int_x \int_y  e^{i (p_x \cdot x + p_y \cdot y) } \vev{ [G^2(x)] [G^2(y)] [G^2(0)] }_c  \nonumber \\[0.1cm]
  &\;\;=\;&   \frac{1}{\pi^2}\frac{1}{\eps} (p_x^{4-2\eps}+p_y^{4-2\eps}+ p_z^{4-2\eps})+ \frac{1}{\pi^2}\left(-\frac{1}{2} \lambda_3 - P_3 \right) + {\cal O}(\eps, g^2) \;, 
 \end{eqnarray}
 where the $2$- and $3$-point kinematic structures $P_3$ and $\la_3$ are defined 
  in \eqref{eq:kine}.  It is observed that at leading order in the coupling there is no divergent contribution to the 
  $3$-point function kinematic structure $\la_3$; or more precisely to the projector  
  $\hat{P}_{\la_3}$  \eqref{eq:3ptProjector} applied to the correlation function.
  The $2$-point function structure $P_3$ 
 is not relevant for our work.
 From \eqref{eq:split} and the definition of $\zamR{gg}{g}{}$ \eqref{eq:ChiABC}, 
 $\chi_{ggg} ={\cal O}(g^2)$ follows.
  In principle this completes the
  task of this section but we  think it is instructive to add a few more comments. 
  
  First, 
  as demonstrated in the appendix \ref{app:formal} 
  this can also be understood from the fact that for an exactly marginal operator the $\la_3$-structure 
  vanishes in a CFT \cite{3scalarCFT}.  In the language of \cite{3scalarCFT} the structure $P_3$ is 
  referred to as semilocal, 
 and that is at least one delta function in coordinate space; 
  $ \de(x)\frac{1}{y^{2d}}   + \textrm{permutations}  \; \leftrightarrow \; p^{4-2 \eps} \frac{1}{\eps}  + \textrm{permutations} $, in the case \eqref{eq:GGG}. 
 For non-coincident points the correlation function is indeed 
proportional to $(d-4)$ cf. \cite{SNR11,DO03}.
   It is in particular instructive to consider a case where 
  this fails. An example is a free conformally coupled scalar field for which $\phi^4$ is an operator of scaling dimension 
  four but since its perturbation $\de {\cal L} \sim \la \phi^4$  induces an RG-flow, namely  $\be_\la \neq 0$, 
  it is not exactly marginal. In the explicit computation, one obtains in coordinate space
\begin{equation}
\label{eq:phiphiphi}
\vev{ [\phi^4(x)] [\phi^4(y)] [\phi^4(0)] }_c = \frac{8}{x^4 y^4 (x-y)^4} \;,
\end{equation} 
which is clearly not semi-local and will contribute to the $\la_3$ structure upon Fourier transformation. 
On a side note it is a remarkable circumstance that from the evaluation 
\eqref{eq:phiphiphi}, one can infer that the $\be_\la \neq 0$, cf. discussion in appendix \ref{app:formal} and 
\cite{3scalarCFT}.

Second, one might wonder whether something similar is possible for  the $2$-point function.  
The answer is no for the following reasons.
If it were possible to set $\zam{A}{B} =0$ in some scheme then 
 it would also imply that $\Theta = 0$ by reflection positivity in \eqref{eq:Dbarb} 
which is incompatible with a non-trivial flow.  The explicit straightforward computation 
for the $2$-point function at leading order gives, 
 $\vev{[G^2(x)] [G^2(0)]}  = 96/x^8 + {\cal O}(g^2)$, a  non-contact term contribution, unlike   \eqref{eq:GGG}, 
 whose Fourier transform gives rise to $\ln \mu$-dependent term. Moreover, the formal argument given in appendix
\ref{app:formal} does not descend to $2$-point functions.  

\subsection{Finiteness of the 3-point Function}
\label{sec:finite3}

Following the analysis in \cite{PZfinite}, we study the finiteness of  $\lDt{g}$, the resummed Laurent series, 
after removing the regulator $\eps = (d-4)/2$. This serves as the basis for defining the $\chischeme$-scheme in the next section. 
The quantity $\lDt{g}$ is defined in analogy to $\lDt{\theta}$ in \eqref{eq:split}.
In dimensional regularisation, the RGE \eqref{eq:ChiABC} reads\footnote{The quantity $\chi_{ggg}^\RR$ 
corresponds to $\bar{\chi}^a_{ggg}$ in the classic paper of Jack and Osborn \cite{JO90}.}
\begin{eqnarray}
\label{eq:RGEggg}
\chi_{ggg}^\RR = (2 \eps - \mathcal{L}_{\be})  \lDt{g}= \left(-2\hat{\be} \partial_{ \ln \als}-6(  \partial_{\ln \als} \hat{\be}) + 2 \eps \right )\lDt{g} \; ,
\end{eqnarray}
where we used the $d$-dimensional logarithmic beta function $\hat{\be}= - \eps + \be $ 
and $a_s \equiv g^2/(4 \pi)^2$.
This equation allows for the integral solution
\begin{eqnarray}
\label{eq:3ptSol1}
\lDt{g}(\als(\mu),\eps)|_{\UV}=  \frac{\mu^{2 \eps}}{\hat{\be}^3} \int_{\ln \mu}^\infty \hat{\be}^{3}(\mu') \chi_{ggg}^\RR(\mu') \frac{d \ln \mu'}{ \mu'^{2 \eps}}  \; .
\end{eqnarray}
This expression is well-defined for $\mu>0$, convergent for $\mu' \to \infty$, as we will show shortly 
for the asymptotically free and asymptotically safe case.
Anticipating the result and removing regulator ($\eps \to 0$)   the expression becomes 
\begin{eqnarray}
\label{eq:3ptSol2}
\lDt{g}(\als(\mu),0)=  \frac{1}{\be^3} \int_{\ln \mu}^\infty \be^{3}(\mu') \chi_{ggg}^\RR d \ln \mu'= -\frac{1}{2 \be^3} \int_0^{\als(\mu)} \be^2(u)\chi_{ggg}^\RR(u)\frac{du}{u} \; ,
\end{eqnarray}
where the second integral representation in \eqref{eq:3ptSol2} is useful for practical computation.

\begin{itemize}
\item In the asymptotically free case (Gaussian FP), 
the $\be$-function and the $3$-metric behave as
\begin{eqnarray} \label{eq:AFfinite}
\be   \sim \als \stackrel{\mu \to \infty}{\sim}   \frac{1}{\ln \mu} \;, \quad \chi_{ggg}^\RR \sim \als \sim \frac{1}{\ln \mu} \;,
\end{eqnarray}
in accordance with \eqref{eq:chi23}. 
Inserting this back into the integral \eqref{eq:3ptSol2} we see that the solution behaves regularly near the UVFP
\begin{eqnarray} \label{eq:ASfinite}
\lDt{g} \stackrel{\mu \to \infty}{\sim} \mathcal{O}(1)   \; ,
\end{eqnarray}
since the $\ln \mu^{-3}$ from the integral is cancelled by the $1/\be^3$ prefactor.
\item
We can also apply similar arguments for RG flows in the vicinity of a non-trivial UVFP $\alsUV$, which corresponds to the asymptotically safe scenario (see \cite{Antipin:2017ebo} for some recent discussion of this possibility). In this case the UV behaviour is 
\begin{equation}
\be \stackrel{\mu \to \infty}{\sim} \mu^{-\gamma^*} \; , \quad \chi_{ggg}^\RR \stackrel{\mu \to \infty}{\sim}
\chi_{ggg}^*  \;,
\end{equation}
where $\chi_{ggg}^*  \equiv \chi_{ggg}^\RR (\alsUV)$ and   $\gamma^* =  \partial_{\ln \als}   \be |_{\als^*} >0$ is the anomalous dimension of $[G^2]$ at the FP. One gets
\begin{equation}
\int_{\ln \mu}^{\infty} \be^3 (\mu') \chi_{ggg}^\RR(\mu') d \ln \mu' \stackrel{\mu \to \infty}{\sim} \frac{1}{3 \gamma^*}\chi_{ggg}^* \mu^{-3 \gamma^*} \; .
\end{equation} 
By inserting this back to \eqref{eq:3ptSol2}, we find again finite UV behaviour 
$\lDt{g} \stackrel{\mu \to \infty}{\sim} \mathcal{O}(1)$. In fact, it is straightforward to see that provided $\gamma^* \neq 0$, for any FP $\als^*$ the equation \eqref{eq:RGEggg} always allows for a finite solution 
\begin{equation} \label{eq:ScalSol}
\lDt{g}(\als^*,0)=- \frac{1}{6 \gamma^*} \chi_{ggg}^* \; .
\end{equation}
\end{itemize}
Let us now turn to the issue of IR convergence. Clearly, the presence of $\frac{1}{\be^3}$ in the solution \eqref{eq:3ptSol2} indicates additional problems when the IR limit $\als \to \alsIR$ is taken. Note however, that near the IRFP another solution to \eqref{eq:RGEggg} can be found
\begin{eqnarray} \label{eq:3ptSolIR2}
\lDt{g}(\als(\mu),\eps)|_{\IR} = - \frac{\mu^{2 \eps}}{\hat{\be}^3} \int_{-\infty}^{\ln \mu} \hat{\be}^{3}(\mu') \chi_{ggg}^\RR(\mu')   \frac{d \ln \mu'} {(\mu')^{2 \eps}} \; ,
\end{eqnarray}
well-defined for $\mu < \infty$.
 After taking $\eps \to 0$ (which is justified for the same reasons as \eqref{eq:3ptSol1}) one gets
\begin{eqnarray}
\label{eq:3ptSolI2}
\lDt{g}(\als(\mu),0)= - \frac{1}{\be^3} \int_{-\infty}^{\ln \mu}  \be^{3}(\mu') \chi_{ggg}^\RR(\mu') d \ln \mu'= -\frac{1}{2 \be^3} \int_\alsIR^{\als(\mu)} \be^2(u)\chi_{ggg}^\RR(u)\frac{du}{u} \; .
\end{eqnarray}
By repeating the analysis leading to \eqref{eq:AFfinite} and \eqref{eq:ASfinite} near the IRFP, we conclude that \eqref{eq:3ptSolIR2} is well-defined in the vicinity of $\alsIR$. 

Assuming that  the theory is free from  any singularities in the coupling space, the solutions \eqref{eq:3ptSol1} and \eqref{eq:3ptSolIR2} should be compatible on overlapping domains. By subtracting  \eqref{eq:3ptSol1} from \eqref{eq:3ptSolIR2} (and taking $\eps \to 0$ limit) we find a continuity condition
\begin{equation} \label{eq:ConditionContinuity}
 \left(\int_{-\infty}^{\ln \mu}+  \int_{\ln \mu}^{\infty}\right)\be^{3}(\mu') \chi_{ggg}^\RR d \ln \mu' =  \int_{-\infty}^\infty  \be^{3}(\mu') \chi_{ggg}^\RR d \ln \mu' = 0 \; .
\end{equation}
This is consistent with the vanishing of the 3-point contribution in \eqref{eq:DelBa1}, which can therefore be seen as the direct consequence of the finiteness and coupling continuity of $\lDt{g}$. Indeed, in the next section, we will show how the above results can be used to construct a scheme, where the 3-metric vanishes.

\subsection{Constructing the $\chischeme$-scheme for the $3$-metric $\chi_{ggg}$}
\label{sec:scheme}

A  change of scheme, cf.~eq.\eqref{eq:Zam3scheme},  is given by a finite shift $\omega_{ggg}(\als)$ in the counterterms 
\begin{equation}
\lDtR{g}{\Rb} = \lDtR{g}{\Ra}- \omega_{ggg}(\als) \;.
\end{equation}
 Using \eqref{eq:RGEggg} we can deduce that under such a  shift the $3$-metric transforms as
 \begin{equation}
 \label{eq:scheme-changeXggg}
   \zamR{g}{gg}{\Rb}  = \zamR{g}{gg}{\Ra} +  \left(2\be \partial_{\ln \als} + 6 (\partial_{\ln \als} \be) \right)\omega_{ggg} \;.
 \end{equation} 
Since the $\eps \to 0$ limit of $\lDtR{g}{\Ra}$ is uniform as shown in the previous section we can choose
\begin{equation}
 \label{eq:3schemeDef}
\omega_{ggg}(\als) \equiv  \lDt{g}(\als, \eps=0) \;,
\end{equation}
 to define a new scheme $\chischeme$ for which 
\begin{eqnarray}
\label{eq:zero}
\zamR{g}{gg}{\chischeme}(\mu)=0 \; ,
\end{eqnarray}
is automatic. This scheme is new to this paper and not to be confused with the previously discussed MOM-scheme.

By using \eqref{eq:zero} in the general scheme-independent expression \eqref{eq:DelBa2} we finally arrive at the desired result
\begin{eqnarray} \label{eq:DelbaQCD}
\Delta \ba = \frac{1}{16}\int_{-\infty}^\infty\zamR{g}{gg}{\RR} \be^2 d \ln \mu' \; .
\end{eqnarray}
We would like to end this section by demonstrating how to construct such a scheme in perturbation theory. Using the two-loop formulas of \cite{JO90}, we extract
\begin{eqnarray} \label{eq:XgggMS}
\zamR{g}{gg}{\MS}= \frac{n_g}{4 \pi^2} (- 2 \be_0 \als) \;  .
\end{eqnarray}
where $\be_0$ is one-loop coefficient of the beta function $\be= - \be_0 \als + \mathcal{O}(\als^2)$, 
the gluons and $N_f$ fermions   are assumed to be in the  adjoint and fundamental representation of an $\SU(N_c)$ gauge group respectively ($n_g \equiv N_c^2-1$),
 and $\MS$ denotes the standard minimal subtraction scheme.
By performing a scheme change \eqref{eq:scheme-changeXggg} with 
\begin{equation} \label{eq:MsChischeme}
\omega_{ggg}= -\frac{1}{3}\frac{n_g}{4 \pi^2} \;,
\end{equation}
 we achieve
\begin{eqnarray}
\zamR{g}{gg}{\chischeme}=0 + \mathcal{O}(\als^2) \;,
\end{eqnarray}
as expected at this order in the perturbation theory.
At the  perturbative Banks-Zaks FP  $\als^\IR \propto \be_0 \ll 1 $. 
Since $\chi_{ggg}$ is absent,  $\Delta \ba$ in this theory can be computed purely by substituting the known perturbative expressions for the beta function and $\chi_{gg}$ in \eqref{eq:Dbarb} as was done up to five loops in \cite{PZboxR} (see Eq. 60 and the discussion bellow in this reference). 

 To find $\zamR{g}{gg}{\chischeme}$ at higher orders one would need to use the two-loop beta function together with the three-loop expression of $\chi_{ggg}$, which is not presently available to the authors. Nevertheless, using the formulae of this paper some general predictions about the behaviour of these higher order corrections can be made (cf. appendix \ref{app:C}). 
 
 After completion of the manuscript $\chi_{ggg}^{\MS}$ was computed to leading order in the large $N_f$ expansion  in \cite{Antipin:2018brm} by resumming infinite number of bubble diagrams. The result reads
\begin{equation}\label{eq:largeNf1}
\chi_{ggg}^\MS = \frac{n_g}{4 \pi^2} \frac{1}{3 K} \frac{\partial}{\partial K} \left( K^2 \bar{H}(\frac{2}{3}K) \right) + \mathcal{O}\left(\frac{1}{N_f}\right)\; ,
\end{equation}
where $K=2 \als N_f$ and
\begin{equation}
\bar{H}(x)= \frac{(80-60x+13x^2-x^3)x \Gamma(4-x)}{120(4-x)\Gamma(1+\frac{x}{2})\Gamma(2-\frac{x}{2})} \;.
\end{equation}
Since to leading order in $\frac{1}{N_f}$ the $\be$-function is one-loop exact $\be = \frac{1}{3}K (1+ O(1/N_f))$, 
it is possible to use our formula \eqref{eq:3ptSol2} together with \eqref{eq:3schemeDef} to find the $\chischeme$ transformation corresponding to \eqref{eq:largeNf1}. The result reads
\begin{equation} \label{eq:largeNf2}
\omega_{ggg}=-\frac{1}{2}\frac{n_g}{4 \pi^2} \frac{\bar{H}(\frac{2}{3}K)}{K}+ \mathcal{O}\left(\frac{1}{N_f}\right)\; .
\end{equation}
The formula above represents an application of our result beyond perturbation theory. By reexpanding expression \eqref{eq:largeNf2} in small $K$ we find that the leading term agrees with \eqref{eq:MsChischeme}.

\section{Conclusions and Discussions}

Our starting point was the   derivation of a formula for the Euler anomaly or $a$-function 
as an integral over the RG-scale of a $2$- and $3$-point functions of the
trace of the energy momentum tensor \eqref{eq:DelBa1}, valid for theories  which are governed by $\be$-functions 
at both fixed points  \eqref{eq:TraceTen}. 

For  gauge theories in the conformal window the formula collapses 
to the $2$-point function \eqref{eq:DelBaProved}. 
Our main assumption for the proof  is that the ultraviolet  and infrared solution \eqref{eq:3ptSol1} and \eqref{eq:3ptSolIR2}, of the RGE 
\eqref{eq:RGEggg}, can be matched continuously. 
This allowed us to define an explicit prescription, the $\chischeme$-scheme  \eqref{eq:3schemeDef},
for which  the $3$-metric vanishes.
 In particular, our result means that for those theories, the Euler flow and the $\Box R$-flow are identical \eqref{eq:Dbarb}. 
The reason this works for gauge couplings, and not for generic couplings, is that 
for the former the $3$-point function collapses to a $2$-point function near the trivial FP \eqref{eq:GGG}. 
This is a consequence of the vanishing of the leading order contribution to the $3$-metric  \eqref{eq:chi23} and can also be understood from the fact that the field strength tensor squared 
is an exactly marginal operator at zero coupling cf. appendix \ref{app:formal}. An example where this fails is 
a scalar free field theory 
for which $\phi^4$ is not an exactly marginal operator and its non-zero $\be$-function  induces a $3$-point function structure at leading order,
prohibiting the use of the $\chischeme$-scheme.

In the light of the above remarks one might wonder, whether the result can be applied to gauge theories 
with supersymmetry including scalar fields such as supersymmetric QCD (SQCD). 
The extension is possible owing the same form of the anomaly \eqref{eq:TraceTen} in SQCD \cite{SV86}. 
In supersymmetric gauge theories without superpotential the matter and gauge contributions are related through the Konishi anomaly \cite{K83}, so that the trace anomaly can be expressed solely in terms of the field 
strength tensor squared up to equation of motion terms. 
We therefore expect that the main results of this paper should apply to $\mathcal{N}=1$ SQCD in the conformal window.\footnote{Some care has to be taken when passing from the dimensional regularisation to SUSY-preserving schemes (see Appendix A of \cite{FO98} for some details of how this is to be done).} 
 
\begin{figure}[t]
\centering
\includegraphics[scale=0.50]{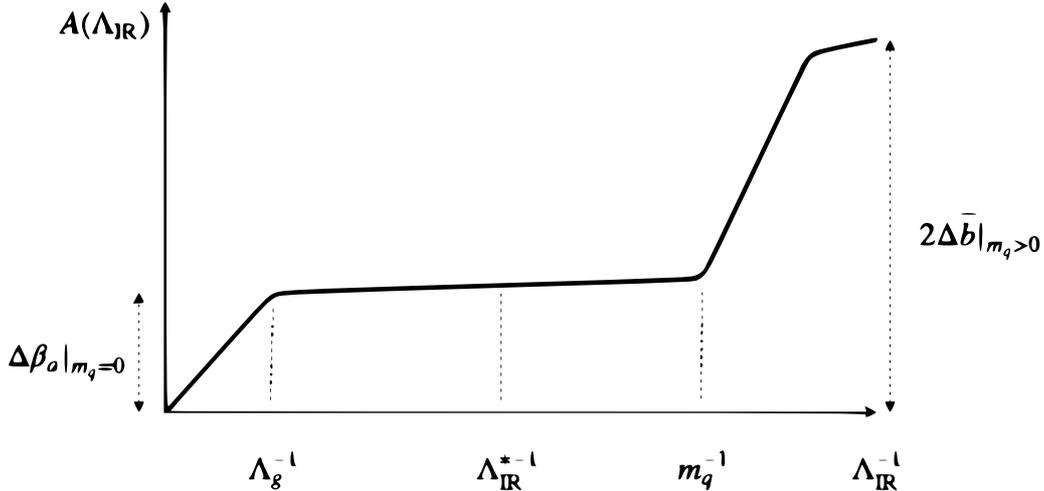}
\caption{\small The Euler flow, $\Delta \be_a$ \eqref{eq:DelBaProved}, as a function of the IR cut-off
$\Lambda_{\IR}$ on the spacial integral \eqref{eq:newA}.  The scale $\Lambda_g^{-1}$ is related to the running of the $\be$-function 
 and can be expected to be of the order of the scale where the derivative of the $\be$-function changes  sign \cite{Pelaggi:2017wzr, Litim:2015iea}. 
 The proposed formula is given in eq.~\eqref{eq:proposed}.
 Note that the asymptotic value $2 \Delta \bar b|_{m_q>0}$ is a non-trivial quantity whose value is not yet understood \cite{PZboxR}. The determination of the latter would therefore be an additional benefit of a lattice investigation}
\label{fig:sketch}
\end{figure} 
 
Another corollary of our analysis is that for the class of theories studied in this paper the  
strong $a$-theorem applies. 
One defines the off-critical quantity
\begin{equation}
    \tilde{\ba}(\mu)= \ba^\UV-\frac{1}{16}\int_{\ln \mu}^\infty\zamS{g}{g} \be^2 d \ln \mu' \; ,
\end{equation}
which reduces to $\ba^\IR$ in the limit $\mu \to 0$ by \eqref{eq:DelbaQCD} and gives a (scheme dependent) interpolating function between the fixed points. The monotonicity of this function follows from the positivity of 
$\zamS{g}{g}$, established in \cite{PZboxR}.  

Moreover, the perspective of implementing the $a$-theorem in the conformal window on the lattice  have improved
 since it is related to a $2$-point function.\footnote{This requires  the renormalisation of 
 the energy momentum tensor on the lattice which is a non-trivial task because of the 
 breaking of the space-time symmetries \cite{Caracciolo:1989pt}. Cf. \cite{Suzuki:2013gza,DelDebbio:2013zaa} for some recent proposals using the gradient flow technique.}  Supersymmetric lattice  gauge theories \cite{Catterall:2009it}, could be a particularly interesting test ground as the Euler anomaly
is exactly known \cite{AFGJ97}.
  In practice though, lattice Monte Carlo simulation are done at finite 
 quark mass which does not fall into 
 our class of theories.\footnote{The TEMT contains a term of a the form 
 $\TEMTO \sim m(1+ \ga_m) \bar qq$ in addition to the $\be$-function terms \eqref{eq:TraceTen} 
 where $\ga_m$ is the quark mass anomalous dimension. 
 Thus unless $\ga_m^\IR = -1$ this does not correspond to a CFT in the IR.} 
 A pragmatic way to deal with this problem is to choose an infrared cutoff $\Lambda_{\IR}^{-1}$,
 \begin{equation}
 \label{eq:newA}
 A(\Lambda_{\IR},m_q,L) 
 \equiv \frac{1}{3 \cdot 2^8}    \int^{\Lambda_\IR^{-1}}_0 d^4 x \, x^4 \vev{\TEMTO(x) \TEMTO(0)}_c
 \end{equation}
  on the 
 integral \eqref{eq:DelBaProved}. The function  $A$ is expected to plateau to $\Delta \be_a|_{m_q =0}$ 
 for  $\Lambda_\IR^{-1}$ lower than the inverse quark mass $m_q$ for which the theory behaves like a massless theory.\footnote{None of this scales should be confused with the lattice size $L$. In particular, 
 $\Lambda_{\IR}^{-1} < L$, holds strictly by construction. To avoid finite size effects one has to impose  
 $m_q^{-1} \ll L$ as 
  for the $2$-point function 
 the quark mass correction ought  to be exponential $\exp(- m_H L)$ where
 $m_H \sim (m_q)^{\eta_H}$ with $\eta_H \equiv 1/(1+\ga^{\IR}_m)$ is a mass 
 of a hadron \cite{DelDebbio:2010jy}.}
More precisely the flat region corresponds to the near-conformal behaviour in the vicinity of the IRFP. Thus one would expect $\Delta \be_a$ to plateau to the massless case 
 \begin{equation}
\label{eq:proposed}
\Delta \be_a|_{m_q =0} \simeq A(\Lambda^*_{\IR},m_q) \;, \qquad   \Lambda_g^{-1} \ll (\Lambda^*_{\IR})^{-1} \ll m_q^{-1} 
\;,
\end{equation} 
 for the above mentioned range, cf. fig.\ref{fig:sketch} for a schematic illustration 
 and an explanation about the scale $\Lambda_g$.
 It would be interesting to  apply this procedure to the case where the IR phase is chirally broken 
 and investigate the expected  appearance of the $\ln m_q$-divergence  induced by the goldstone bosons \cite{LPR12,PZboxR}.

\appendix
\numberwithin{equation}{section}

\subsection*{Acknowledgements}
 
We are grateful to Luigi Del Debbio, Jos\'e Latorre  and Kostas Skenderis  for useful discussions and
Saad Nabeebaccus for careful reading of the manuscript. 
VP acknowledges the support of the Higgs Centre Edinburgh 
where a large part of this work was completed. VP is supported by the ERC STG grant 639220 (curvedsusy).

\section{Conventions for computing correlators}

\label{app:Conventions}

In this appendix we will give a brief review of the notation related to the renormalization of composite operator correlators adapted from \cite{PZfinite}. 

We start with the $2$-point functions of classically marginal renormalised operators $[O_A]$
\begin{eqnarray}
\label{eq:GAB}
\Gamma_{AB}(p^2)   =   
 \int d^4 x  e^{i p \cdot x} \vev{[O_A(x)][ O_B(0)]}_c   =
  \Gamma_{AB }^\RR(p^2, \mu)   + \lnDR{A}{B}{\RR}  \mu^{-2\eps}  p^4 \;,
\end{eqnarray}
where the subtraction constant $\lnDR{A}{B}{\RR}$ is a function of couplings of the theory and it contains Laurent series in $\eps$ as well as a finite part and $\Gamma_{AB }^\RR(p^2 , \mu)$ is the finite renormalized correlator. A scheme $\RR$ is determined by the choice of the finite part of $\lnDR{A}{B}{\RR}$. \\
The finite quantity called Zamolodchikov metric is obtained via
\begin{equation}
\label{eq:RGEm}
(    2 \eps-{\cal L}_\be )  \lnDR{A}{B}{\RR} =   \zamR{A}{B}{\RR} \;.
\end{equation}
where ${\cal L}_\be$ denotes the Lie derivative on a $2$ tensor in coupling space
\begin{equation}
{\cal L}_\be  \lnDR{A}{B}{\RR}  = 
\partial_A \hat{\be}^C  \lnDR{C}{B}{\RR}+\partial_B \hat{\be}^C  \lnDR{A}{C} {\RR}+ 
 \hat{\be}^C \partial_C \lnDR{A}{B}{\RR} \;.
\end{equation}
Since the bare correlator is scale independent, 
the Zamolodchikov metric can be also defined directly from the finite renormalized correlators, 
e.g. \cite{Shore:2016xor}, as follows
\begin{eqnarray}\label{eq:FiniteCorrMetric}
(-\frac{\partial}{\partial \ln \mu}+ {\cal L}_\be)\Gamma_{AB }^\RR(p^2, \mu)= \zamR{A}{B}{\RR} p^4\;.
\end{eqnarray}
A scheme change is implemented using a finite, coupling dependent constant $\omega_{AB}$ via 
\begin{equation}
\Gamma_{AB }^\RR \to \Gamma_{AB }^\RR + \omega_{AB} p^4\;, \quad 
\lnDR{A}{B}{\RR} \to \lnDR{A}{B}{\RR} - \omega_{AB} \;,
\end{equation}
 leaving the bare correlator  invariant. Under such change we get a shift in the Zamolodchikov metric
\begin{eqnarray} \label{eq:Zam2scheme}
\zamR{A}{B}{\RR} \to \zamR{A}{B}{\RR} + {\cal L}_\be \omega_{AB} \;.
\end{eqnarray}

The $3$-point functions can be defined in a similar manner
\begin{eqnarray}
\label{eq:GABC}
\Gamma_{ABC}(p_x,p_y)  &=&   
 \int d^4 x  d^4 y  e^{i (p_x \cdot x + p_y \cdot y) } \vev{[O_A(x)][ O_B(y)][ O_C(0)]}_c   \nonumber  \\[0.1cm]
   &=& \Gamma_{ABC}^\RR +  \lnD{(A)}{BC}  p_x^4   + \lnD{A}{BC}   P_{yz} +  \text{cyclic} \;,
 \end{eqnarray}
where cyclic permutation over the pairs $(A,x)$, $(B,y)$ and $(C,z)$ is implied. 
Furthermore,  $p_x + p_y + p_z =0$, and $P_{yz} = p_x^4 - p_x^2(p_y^2+p_z^2)$ are kinematic structures vanishing whenever \emph{any} of the three external momenta $p_{x,y,z}$ is set to zero. Just as above $\lnD{(A)}{BC}$ and $\lnD{A}{BC}$ are subtraction constants containing Laurent series and finite parts.
It follows that the $\lnD{(A)}{BC}$-coefficients  can be determined from the 2-point functions information (see \cite{PZfinite} for the exact definition).

The new, purely 3-point information is encoded in the $ \lnDR{A}{BC}{\RR}$ tensor. Again the scale 
derivative  
\begin{equation}
\label{eq:ChiABC}
   (  2 \eps-{\cal L}_{\be} )   \lnDR{A}{BC}{\RR} =   
\zamR{A}{BC}{\RR} \;,  
\end{equation}
proves useful. Above  ${\cal L}_{\be}$  denotes the Lie derivative, acting on a $3$-tensor 
\begin{equation}
{\cal L}_{\be} \lnDR{A}{BC}{\RR}  =   \partial_A \hat{\be}^D  \lnDR{D}{BC}{\RR}+\partial_B \hat{\be}^D     \lnDR{A}{DC}{\RR} +\partial_C \hat{\be}^D  \lnDR{A}{BD}{\RR} +  \hat{\be}^D \partial_D   \lnDR{A}{BC}{\RR} \;.
\end{equation}
We can also define $\zamR{A}{BC}{\RR}$ through the finite renormalized correlator $\Gamma_{ABC}^\RR$ by using a relation analogical to \eqref{eq:FiniteCorrMetric} and projecting onto $P_{yz}$ etc. \\
The scheme change in $\chi_{ABC}^\RR$ is implemented via constant $\omega_{ABC}$, which is now \textit{independent} of $\omega_{AB}$ from \eqref{eq:Zam2scheme} (this follows since the structure $\lnD{A}{BC}$ is independent of the $2$-point function). Under such scheme change we have $\lnD{A}{BC} \to \lnD{A}{BC} - \omega_{ABC}$ and therefore
\begin{eqnarray}
 \label{eq:Zam3scheme}
\zamR{A}{BC}{\RR} \to \zamR{A}{BC}{\RR} + {\cal L}_{\be} \omega_{ABC} \;.
\end{eqnarray}

\section{Derivation of the 3-point Sum Rule}
\label{app:sumrules}

In this appendix, we provide a derivation of the $3$-point sum rule \eqref{eq:3ptSumRuleBa} used 
in section  \ref{sec:sumrules}  to derive a RG-scale integral representation 
for the Euler flow $\Delta \ba$.  Using the quantum action principle, a constraint
on a gravity counterterm is worked out in section \ref{app:La}, which is then used 
in the anomaly matching argument in section \ref{app:DilatSumRule} to derive the sum rule.

\subsection{Renormalisation in Curved Space and  $\ba^\UV$}
\label{app:La}

In an external gravitational field one needs to add  counterterms
\begin{equation}
 \label{eq:Lgrav}
 {\cal L}_{\textrm{gravity}} = -( a_0 E_4 + c_0 W^2 +   b_0  H^2)   \;,
 \end{equation}
 to the action to renormalise the theory \cite{H81}.
The bare couplings are defined as $a_0 = \mu^{d-4} (a^\RR(\mu) + L_a^\RR(\mu))$  etc. and geometric  quantities are the same as in \eqref{eq:VEVTEMT} except for $\Box R$ which being a total derivative does not contribute to the action. 
The main idea is that the quantum action principle (differentiation with respect to sources) 
leads to finite quantities and thus to constraints on the counterterms.
Concretely, a triple Weyl variation $\de s(x)$  ($\gm_{\mu\nu} \to e^{ - 2 s(x)} \gm_{\mu\nu}$) leads to
\begin{eqnarray}
\label{eq:3point}
 \int_{x} \int_y  e^{i (p_x \cdot x + p_y \cdot y) }\frac{\delta^3}{\delta s(x)\delta s(y)\delta s(0)}\ln \Zpart  = 
  ( 2\kd a_0-8b_0)\la_3   + \Gamma_{\theta \theta \theta}(p_x,p_y) =  \fin   \;,
\end{eqnarray}
where the abbreviation $\kd \equiv (d-4)(d-3)(d-2)$ is introduced and we used \eqref{eq:FTttt} to include the dynamical contribution.  By using, \eqref{eq:split} we conclude that the finiteness of $\lnD{\theta} {\theta \theta} $ ensures finiteness of the combination $(2\kd a_0-8b_0)$ in \eqref{eq:3point}. 
Since $b_0$ has been shown to be finite \cite{PZfinite} it is to be  concluded that the quantity $\kd a_0$ is finite. In particular this means that the $\eps \to 0$ limit  $\kd a_0$ is meaningful
\begin{equation}
\label{eq:FiniteA}
 \lim_{\eps \to 0} \kd a_0  \equiv \lim_{\eps \to 0} \kd (L_a^\UV+ a^\UV)= - 2 \ba^\UV \;.
\end{equation}
In the last step we used that $a^\UV$ is finite and that  $L_a^\UV= \frac{\ba^\UV}{2 \eps}$.
The latter follows from $\ba=  -(\frac{d}{d \ln \mu}  -2 \eps) L_a $ and 
the stationarity property $\frac{d}{d \ln \mu}L_a^\UV = 0$ at FPs (which can be seen 
by writing $ L_a \sim x_1 +  x_2 (g^I-g^{I,\UV}) $ with $x_{1,2}$  constants and  using  $\be^{I,\UV} =0$). 
Eq. \eqref{eq:FiniteA} is a relevant observation as this  
implies finiteness of the corresponding term in the dilaton effective action.

\subsection{Sum Rule from the Dilaton Effective Action}
\label{app:DilatSumRule}

In the $3$-point sum rule \eqref{eq:3ptSumRuleBa}, the Euler flow  $\be_a$  
 arises, in 
dimensional regularisation, from  an evanescent operator.  
This can be seen by writing the  $d$-dimensional Euler term 
 as a sum of a four dimensional  and  an evanescent term 
\begin{equation}
 \label{eq:EulerApp}
\sqrt{\gm} E_d  = \sqrt{\gm} E_4  -\kd e^{2\eps s} (-2 \Box s (\partial s)^2 + (\partial s)^4  - 2 \epsilon (\partial s)^4)    \;,
\end{equation}
where we have assumed the conformally flat metric $\gm_{\al \be} = e^{-2s(x) } \de_{\al \be}$ 
and $\kd \sim \eps $ is defined below \eqref{eq:3point}. 
The $\sqrt{\gm} E_4$-term  is a total derivative 
\begin{equation}
\sqrt{\gm} E_4  = \partial O=-4(d-3)(d-2)  \left[ \frac{1}{2} \Box (e^{2\eps s} (\partial s)^2) + \partial ( e^{2\eps s} \partial s ((1-\eps)(\partial s)^2-\Box s))   \right] \;,
\end{equation}
characteristic of topological terms.
The  evanescent part of the gravitational counterterms \eqref{eq:Lgrav} becomes 
the Wess-Zumino term of the dilaton effective action in \cite{KS11}
\begin{eqnarray}
\label{eq:WZcterm}
 {\cal L}_{\textrm gravity}  &\supset&  a_0 \int d^d x \sqrt{g} (E_d - E_4) = -\kd a_0 \int d^d x (-2 \Box s (\partial s)^2 + (\partial s)^4  - 2 \epsilon (\partial s)^4)  \nonumber \\[0.1cm]
 & \stackrel{\eps \to 0}{\to} & 2 \ba^{\UV}  \int d^4 x (-2 \Box s (\partial s)^2 + (\partial s)^4)= 2 \ba^{\UV} S_{WZ} \;,
\end{eqnarray}
where we have used \eqref{eq:FiniteA}. In the preceding argument, the finiteness of $\kd a_0$ (and $b_0$) was essential to ensure UV finiteness of the dilaton effective action and match the the bare coefficient of the Wess-Zumino term to the Euler anomaly $\ba^{\UV}$. 

Similarly, the   IR effective action contains the term $2 \ba^{\IR} S_{WZ}$ which  contributes at ${\cal O}(s^3)$
\begin{equation}
\label{eq:IReff3s}
\ln \Zpart = -4 \barb^\IR \int_x (\Box s)^2 - (4 \ba^{\IR} - 8\barb^\IR) \int_x (\partial s)^2\Box s \ + \dots \quad \;.
\end{equation}
 We are now ready to put all the pieces together. By Fourier transforming the third functional derivative with respect to $s$ of \eqref{eq:IReff3s}, we see that at low momenta, the LHS of \eqref{eq:3point} behaves as
\begin{equation}
\label{eq:C9}
-(4 \ba^{\IR} - 8\barb^\IR)\lambda_3  + \dots \;,
\end{equation}
where the dots stand for nonlocal contributions subleading in the momentum expansion. 
Assuming momentum conservation, $p_z = - (p_x + p_y)$, $\la_3$ \eqref{eq:kine} assumes the form
\begin{equation}
\lambda_3= 4 \left[(p_x \cdot p_y)^2- p_x^2 p_y^2\right] \;,
\end{equation}
with the associated projector $\hat{P}_{\la_3} \la_3 = 1$ being 
\begin{equation}
\label{eq:3ptProjector}
\hat{P}_{\la_3}=\frac{1}{96} \left[(\partial_{p_x} \cdot \partial_{p_y})^2 - \partial_{p_x}^2 \partial_{p_y}^2  \right] \;,
\end{equation}
for which the $P_3$-structure automatically vanishes ($\hat{P}_{\la_3} P_3 =0$). 
 Applying $ \hat{P}_{\la_3} $ to the right-hand side of \eqref{eq:3point}, one gets
\begin{equation}
- \hat{P}_{\la_3}\Gamma_{\theta \theta \theta}(p_x,p_y) |_{p_x=p_x=0} - (4 \ba^{\UV} - 8\barb^\UV)= -(4 \ba^{\IR} - 8\barb^\IR) \;,
\end{equation}
where we used that $(2\kd a_0-8b_0) \to -(4 \ba^{\UV}- 8\barb^\UV)$ for $\eps \to 0$.
The $3$-point sum rule in momentum space follows
\begin{equation}
\label{eq:refer}
\Delta \ba = 2 \Delta \barb -\frac{1}{4}\hat{P}_{\la_3} \Gamma_{\theta \theta \theta}(p_x,p_y) |_{p_x=p_y=0} \;.
\end{equation}
which in  position space assumes the form 

\begin{eqnarray}
\label{eq:DelaMaster}
\Delta \ba  =  \frac{1}{3 \cdot 2^8}\big(  \underbrace{ \int_x x^4 \vev{\TEMTO(x) \TEMTO(0)}_c}_{3 \cdot  2^9 \cdot \Delta \barb} - 2
\int_{x} \int_y  \ [ (x \cdot y)^2 - x^2y^2] 
\vev{\TEMTO(x)\TEMTO(y)\TEMTO(0)}_c \big) \;.
\end{eqnarray}

The Euler flow formula \eqref{eq:DelaMaster} is invariant under the addition of 
the  local $\de {\cal L} \sim \omega_0 R^2$-term unlike the sum rule for $\Delta \barb$ \eqref{eq:Dbarb} 
which needs to be amended. Such a scheme change (denoted as '$R^2$-scheme' in \cite{PZboxR}) should be viewed as independent of \eqref{eq:Zam2scheme} and \eqref{eq:Zam3scheme}.\footnote{For more thorough discussion of various classes of schemes we refer the reader to \cite{PZboxR}, namely sections 2.3.2 and 2.3.3 of this reference.}
More concretely, the contribution of such shift precisely cancels between $2$- and $3$-point parts in \eqref{eq:DelaMaster}. 
The reason this has to happen is that $\be_a$ is well-defined at each FP and not only as a difference, like 
$\Delta \barb$.
At last, we would like to mention that eq.~\eqref{eq:DelaMaster} itself has been derived 
by Anselmi \cite{A03} using different methods.

\section{Vanishing of $\chi_{ggg}$ at the UV Fixed Point - formal Argument}
\label{app:formal}

In this appendix we will demonstrate how $\chi_{ggg}=0$ in the free theory can be derived by using standard OPE arguments \cite{Kadanoff:1978pv}. We start the discussion by considering a general perturbation 
\begin{equation}
  \de S =   \lambda \int_x O_\lambda(x) \; ,
\end{equation}
for some coupling constant $\lambda$ that can be set to $1$ without loss of generality. We now deform the constant  $\lambda \to \lambda + \delta \lambda$, the corrections to a generic correlator $\vev{\dots}$ in the perturbed theory read
\begin{eqnarray} \label{eq:dlamExp}
    \vev{\dots} = \vev{\dots}_{\delta \lambda=0}  &\;+\;&  \delta \lambda \int_x \vev{O_\lambda(x)\dots}_{\delta \lambda=0}  \nonumber \\[0.1cm]
     &\;+\;& 
     \frac{1}{2}\delta \lambda^2 \int_x \int_y \vev{O_\lambda(x)O_\lambda(y) \dots }_{\delta \lambda=0} + \mathcal{O}(\delta \lambda^3) \;.
\end{eqnarray}
The $\delta \lambda^2$-term in \eqref{eq:dlamExp} can be obtained by using the OPE
\begin{eqnarray}
O_\lambda(x)O_\lambda(y)= \frac{C_{\lambda \lambda}^\la}{|x-y|^4}O_\lambda(x) + \dots \; ,
\end{eqnarray}
where dots encompass terms irrelevant for the calculation. By inserting this expression  back into \eqref{eq:dlamExp} and evaluating the $\int_y$ integral with a UV cutoff $\Lambda$, one finds that a logarithmic divergence appears
\begin{equation} \label{eq:dlamExp1}
\sim  C_{\lambda \lambda}^\lambda \delta \lambda^2 \ln \Lambda  \int_x \vev{O_\lambda(x)\dots}_{\delta \lambda=0}  \; .
\end{equation}
This divergence can be removed by adding a counterterm of the form
\begin{equation}
    \delta \lambda^2 C_{\lambda\lambda}^\lambda \ln{(\Lambda/\mu)} \int_x O_\lambda(x) \; .
\end{equation}
However, adding such a term amounts to introducing a $\be$-function
\begin{equation} \label{eq:LamBeta}
\be_\lambda \sim C_{\lambda \lambda}^\lambda \delta \lambda^2 + \mathcal{O}(\delta \lambda^3) \; .
\end{equation}
Hence the non-vanishing of the $\be_\la$-function and the OPE coefficient  $C_{\lambda \lambda}^\lambda$
are directly related. 

We  restrict our attention to the case at hand where $O_\lambda \equiv O_g= G^2$ in the free-field theory.  Since the value of $\delta g$ only affect the normalisation of the kinetic term, it is clear that the theory remains free (and therefore a CFT) for any value of $\delta g$. Thus $\be =0$ and in the free theory
\begin{eqnarray}
C_{gg}^g = 0 \;.
\end{eqnarray}
 Using the fact  that in a CFT the $3$-point function of an operator $O_\la$ is proportional to $C_{\la\la}^{\la}$ we conclude that also the $3$-point function of $O_g$ has to vanish, which directly implies that
\begin{eqnarray}
\chi_{ggg}^\text{free}=0 \; .
\end{eqnarray}
Note that the above argument shows that $O_g$ is exactly marginal at the Gaussian fixed point. In general an operator is called exactly marginal if its beta function vanishes, which is equivalent to the vanishing of the corresponding 3-point function as shown in \cite{Kadanoff:1978pv,3scalarCFT}.\footnote{Note that the non-zero QCD $\be$-function should be understood as a consequence of coupling to fermions and gluons rather than a deformation by $G^2$.}

\section{The $\chischeme$-scheme in Perturbation Theory}
 \label{app:C}

In this appendix, we will construct explicitly the solutions \eqref{eq:3ptSol2} and \eqref{eq:3ptSolIR2} for 
a theory with a trivial UVFP and a Banks-Zaks FP in the IR. 

We start with the analysis near the trivial UVFP.  
The scheme \eqref{eq:3schemeDef}
 means that given a $\chi_{ggg}^\RR$, we should be able to obtain $\omega_{ggg}$ to any given order in $\als$ through  \eqref{eq:3ptSol2}. We  demonstrate how this works for the first two non-vanishing orders in perturbation theory. 
Introducing the notation 
\begin{eqnarray} \label{eq:3metric2loop}
\chi_{ggg}^\RR= \chi_{ggg}^{(1)}\als + \chi_{ggg}^{(2)}\als^2 + \mathcal{O}(\als^3) \; ,
\end{eqnarray}
the scheme change $\omega_{ggg}$ to $\mathcal{O}(\als^2)$ is given by, 
performing the integral on the right-hand side of \eqref{eq:3ptSol2} and expanding the result in $\als$, 
\begin{eqnarray} 
\label{eq:wPert1}
\omega_{ggg}\|_{\UV}= \frac{\chi_{ggg}^{(1)}}{6\be_0}+ \frac{\left(2 \be_1 \chi_{ggg}^{(1)}- \be_0 \chi_{ggg}^{(2)}\right)\als}{2\be_0^2}-  
\frac{7\left(2\be_1^2 \chi_{ggg}^{(1)}- \be_0 \be_1 \chi_{ggg}^{(2)} \right)\als^2}{4 \be_0^3} + \mathcal{O}(\als^3) \; ,
\end{eqnarray}
where the two-loop $\be$-function is parameterised by
$\be=-\be_0 \als - \be_1 \als^2$.
 It is easily verified that
 \begin{equation} \label{eq:schemePertchange}
   \zamR{g}{gg}{\chischeme}  = \zamR{g}{gg}{\RR} +  \left(2\be \partial_{\ln \als} + 6 (\partial_{\ln \als} \be) \right)\omega_{ggg} = 0 + \mathcal{O}(\als^3)\;.
 \end{equation} 
 Note that \eqref{eq:wPert1} is $\mathcal{O}(1)$ and therefore nonzero at the UVFP even though $\chi_{ggg}^\RR$ itself  vanishes there. 
 
In the  IR we assume a (Banks-Zaks) FP at $\alsIR= -\frac{\be_0}{\be_1} \ll 1$, which exists for asymptotically free theory with $\be_0>0, \be_1<0$. 
The $3$-metric \eqref{eq:3metric2loop} expands to 
 \begin{equation}
 \chi_{ggg}^{*} \equiv \zamR{g}{gg}{\RR}(\alsIR)= \frac{\be_0^2}{\be_1}(-r+\frac{\chi_{ggg}^{(2)}}{\be_1}) + \mathcal{O}(\be_0^3) \; ,
\end{equation}  
where we used that $\chi_{ggg}^{(1)}= \be_0 r$ for some finite constant $r$ (c.f. \eqref{eq:XgggMS}).
 Close to this FP,  \eqref{eq:3ptSolIR2} admits a perturbative solution in $\Delta \als \equiv \als - \alsIR$ 
 \begin{equation} \label{eq:wPert2}
 \omega_{ggg}|_{\IR} =\frac{\chi_{ggg}^{*}}{6 \gamma^*}- \frac{(2 \be_1 r-  \chi_{ggg}^{(2)})}{4 \be_0}\Delta  \als + \frac{7(2 \be_1^2 r -  \be_1 \chi_{ggg}^{(2)})}{20 \be_0^2} (\Delta \als)^2 + \mathcal{O}( (\Delta \als)^3) \; .
 \end{equation}
Above we used that  $\gamma^* =-
 \partial_{\ln \als} \be|_{\als^*} = 
 \frac{\be_0^2}{\be_1}$ is the  anomalous dimension of $O_g$ (field strength tensor squared). 
In the limit $\als \to \alsIR$ limit we get the expected dependence \eqref{eq:ScalSol}.
 Direct computation shows that \eqref{eq:wPert2} is compatible with  \eqref{eq:schemePertchange}. The leading $\mathcal{O}(1)$ parts of the IR solution \eqref{eq:wPert2} and the UV solution \eqref{eq:wPert1} match, up to $\mathcal{O}(\be_0)$ corrections,  provided that
\begin{equation}
-2\be_1 r + \chi_{ggg}^{(2)}= \mathcal{O}(\be_0) \; .
\end{equation}
The above should be regarded as the necessary condition for continuity of $\omega_{ggg}$, equivalent to \eqref{eq:ConditionContinuity}.

\bibliographystyle{utphys}
\bibliography{input4}

\providecommand{\href}[2]{#2}\begingroup\raggedright\begin{thebibliography}{10}

\bibitem{20years}
M.~Duff, ``{Twenty years of the Weyl anomaly},''
  \href{http://dx.doi.org/10.1088/0264-9381/11/6/004}{{\em Class.Quant.Grav.}
  {\bfseries 11} (1994) 1387--1404},
\href{http://arxiv.org/abs/hep-th/9308075}{{\ttfamily arXiv:hep-th/9308075
  [hep-th]}}.

\bibitem{Zamolodchikov:1986gt}
A.~B. Zamolodchikov, ``{Irreversibility of the Flux of the Renormalization
  Group in a 2D Field Theory},'' {\em JETP Lett.} {\bfseries 43} (1986)
  730--732.
[Pisma Zh. Eksp. Teor. Fiz.43,565(1986)].

\bibitem{Cardy:1988tj}
J.~L. Cardy, ``{The Central Charge and Universal Combinations of Amplitudes in
  Two-dimensional Theories Away From Criticality},''
\href{http://dx.doi.org/10.1103/PhysRevLett.60.2709}{{\em Phys. Rev. Lett.}
  {\bfseries 60} (1988) 2709}.

\bibitem{PZboxR}
V.~Prochazka and R.~Zwicky, ``{On the Flow of $\Box R$ Weyl-Anomaly},''
  \href{http://dx.doi.org/10.1103/PhysRevD.96.045011}{{\em Phys. Rev.}
  {\bfseries D96} no.~4, (2017) 045011},
\href{http://arxiv.org/abs/1703.01239}{{\ttfamily arXiv:1703.01239 [hep-th]}}.

\bibitem{Cardy:1988cwa}
J.~L. Cardy, ``{Is There a c Theorem in Four-Dimensions?},''
\href{http://dx.doi.org/10.1016/0370-2693(88)90054-8}{{\em Phys. Lett.}
  {\bfseries B215} (1988) 749--752}.

\bibitem{KS11}
Z.~Komargodski and A.~Schwimmer, ``{On Renormalization Group Flows in Four
  Dimensions},'' \href{http://dx.doi.org/10.1007/JHEP12(2011)099}{{\em JHEP}
  {\bfseries 12} (2011) 099},
\href{http://arxiv.org/abs/1107.3987}{{\ttfamily arXiv:1107.3987 [hep-th]}}.

\bibitem{K11}
Z.~Komargodski, ``{The Constraints of Conformal Symmetry on RG Flows},''
  \href{http://dx.doi.org/10.1007/JHEP07(2012)069}{{\em JHEP} {\bfseries 07}
  (2012) 069},
\href{http://arxiv.org/abs/1112.4538}{{\ttfamily arXiv:1112.4538 [hep-th]}}.

\bibitem{FL98}
S.~Forte and J.~I. Latorre, ``{A Proof of the irreversibility of
  renormalization group flows in four-dimensions},''
  \href{http://dx.doi.org/10.1016/S0550-3213(98)00631-2}{{\em Nucl. Phys.}
  {\bfseries B535} (1998) 709--728},
\href{http://arxiv.org/abs/hep-th/9805015}{{\ttfamily arXiv:hep-th/9805015
  [hep-th]}}.

\bibitem{Osborn:1989td}
H.~Osborn, ``{Derivation of a Four-dimensional $c$ Theorem},''
\href{http://dx.doi.org/10.1016/0370-2693(89)90729-6}{{\em Phys. Lett.}
  {\bfseries B222} (1989) 97--102}.

\bibitem{BKZR14}
F.~Baume, B.~Keren-Zur, R.~Rattazzi, and L.~Vitale, ``{The local
  Callan-Symanzik equation: structure and applications},''
  \href{http://dx.doi.org/10.1007/JHEP08(2014)152}{{\em JHEP} {\bfseries 08}
  (2014) 152},
\href{http://arxiv.org/abs/1401.5983}{{\ttfamily arXiv:1401.5983 [hep-th]}}.

\bibitem{A03}
D.~Anselmi, ``{Kinematic sum rules for trace anomalies},''
  \href{http://dx.doi.org/10.1088/1126-6708/2001/11/033}{{\em JHEP} {\bfseries
  11} (2001) 033},
\href{http://arxiv.org/abs/hep-th/0107194}{{\ttfamily arXiv:hep-th/0107194
  [hep-th]}}.

\bibitem{Intriligator:1995au}
K.~A. Intriligator and N.~Seiberg, ``{Lectures on supersymmetric gauge theories
  and electric-magnetic duality},''
  \href{http://dx.doi.org/10.1016/0920-5632(95)00626-5}{{\em Nucl. Phys. Proc.
  Suppl.} {\bfseries 45BC} (1996) 1--28},
  \href{http://arxiv.org/abs/hep-th/9509066}{{\ttfamily arXiv:hep-th/9509066
  [hep-th]}}.
[,157(1995)].

\bibitem{Hill:2002ap}
C.~T. Hill and E.~H. Simmons, ``{Strong dynamics and electroweak symmetry
  breaking},'' \href{http://dx.doi.org/10.1016/S0370-1573(03)00140-6}{{\em
  Phys. Rept.} {\bfseries 381} (2003) 235--402},
  \href{http://arxiv.org/abs/hep-ph/0203079}{{\ttfamily arXiv:hep-ph/0203079
  [hep-ph]}}.
[Erratum: Phys. Rept.390,553(2004)].

\bibitem{LS89}
H.~Leutwyler and M.~A. Shifman, ``{GOLDSTONE BOSONS GENERATE PECULIAR CONFORMAL
  ANOMALIES},''
\href{http://dx.doi.org/10.1016/0370-2693(89)91730-9}{{\em Phys. Lett.}
  {\bfseries B221} (1989) 384--388}.

\bibitem{DL91}
J.~F. Donoghue and H.~Leutwyler, ``{Energy and momentum in chiral theories},''
\href{http://dx.doi.org/10.1007/BF01560453}{{\em Z. Phys.} {\bfseries C52}
  (1991) 343--351}.

\bibitem{LPR12}
M.~A. Luty, J.~Polchinski, and R.~Rattazzi, ``{The $a$-theorem and the
  Asymptotics of 4D Quantum Field Theory},''
  \href{http://dx.doi.org/10.1007/JHEP01(2013)152}{{\em JHEP} {\bfseries 01}
  (2013) 152},
\href{http://arxiv.org/abs/1204.5221}{{\ttfamily arXiv:1204.5221 [hep-th]}}.

\bibitem{A99}
D.~Anselmi, ``{Anomalies, unitarity and quantum irreversibility},''
  \href{http://dx.doi.org/10.1006/aphy.1999.5949}{{\em Annals Phys.} {\bfseries
  276} (1999) 361--390},
\href{http://arxiv.org/abs/hep-th/9903059}{{\ttfamily arXiv:hep-th/9903059
  [hep-th]}}.

\bibitem{PZfinite}
V.~Prochazka and R.~Zwicky, ``{Finiteness of two- and three-point functions and
  the renormalization group},''
  \href{http://dx.doi.org/10.1103/PhysRevD.95.065027}{{\em Phys. Rev.}
  {\bfseries D95} no.~6, (2017) 065027},
\href{http://arxiv.org/abs/1611.01367}{{\ttfamily arXiv:1611.01367 [hep-th]}}.

\bibitem{JO90}
I.~Jack and H.~Osborn, ``{Analogs for the $c$ Theorem for Four-dimensional
  Renormalizable Field Theories},''
\href{http://dx.doi.org/10.1016/0550-3213(90)90584-Z}{{\em Nucl.Phys.}
  {\bfseries B343} (1990) 647--688}.

\bibitem{3scalarCFT}
A.~Bzowski, P.~McFadden, and K.~Skenderis, ``{Scalar 3-point functions in CFT:
  renormalisation, beta functions and anomalies},''
  \href{http://dx.doi.org/10.1007/JHEP03(2016)066}{{\em JHEP} {\bfseries 03}
  (2016) 066},
\href{http://arxiv.org/abs/1510.08442}{{\ttfamily arXiv:1510.08442 [hep-th]}}.

\bibitem{SNR11}
S.~El-Showk, Y.~Nakayama, and S.~Rychkov, ``{What Maxwell Theory in D<>4
  teaches us about scale and conformal invariance},''
  \href{http://dx.doi.org/10.1016/j.nuclphysb.2011.03.008}{{\em Nucl. Phys.}
  {\bfseries B848} (2011) 578--593},
\href{http://arxiv.org/abs/1101.5385}{{\ttfamily arXiv:1101.5385 [hep-th]}}.

\bibitem{DO03}
F.~A. Dolan and H.~Osborn, ``{Conformal partial waves and the operator product
  expansion},'' \href{http://dx.doi.org/10.1016/j.nuclphysb.2003.11.016}{{\em
  Nucl. Phys.} {\bfseries B678} (2004) 491--507},
\href{http://arxiv.org/abs/hep-th/0309180}{{\ttfamily arXiv:hep-th/0309180
  [hep-th]}}.

\bibitem{Antipin:2017ebo}
O.~Antipin and F.~Sannino, ``{Conformal Window 2.0: The large $N_f$ safe
  story},'' \href{http://dx.doi.org/10.1103/PhysRevD.97.116007}{{\em Phys.
  Rev.} {\bfseries D97} no.~11, (2018) 116007},
\href{http://arxiv.org/abs/1709.02354}{{\ttfamily arXiv:1709.02354 [hep-ph]}}.

\bibitem{Antipin:2018brm}
O.~Antipin, N.~A. Dondi, F.~Sannino, and A.~E. Thomsen, ``{The $a$-theorem at
  large $N_f$},''
\href{http://arxiv.org/abs/1808.00482}{{\ttfamily arXiv:1808.00482 [hep-th]}}.

\bibitem{SV86}
M.~A. Shifman and A.~I. Vainshtein, ``{Solution of the Anomaly Puzzle in SUSY
  Gauge Theories and the Wilson Operator Expansion},''
  \href{http://dx.doi.org/10.1016/0550-3213(86)90451-7}{{\em Nucl. Phys.}
  {\bfseries B277} (1986) 456}.
[Zh. Eksp. Teor. Fiz.91,723(1986)].

\bibitem{K83}
K.~Konishi, ``{Anomalous Supersymmetry Transformation of Some Composite
  Operators in SQCD},''
\href{http://dx.doi.org/10.1016/0370-2693(84)90311-3}{{\em Phys.Lett.}
  {\bfseries B135} (1984) 439}.

\bibitem{FO98}
D.~Z. Freedman and H.~Osborn, ``{Constructing a c function for SUSY gauge
  theories},'' \href{http://dx.doi.org/10.1016/S0370-2693(98)00649-2}{{\em
  Phys. Lett.} {\bfseries B432} (1998) 353--360},
\href{http://arxiv.org/abs/hep-th/9804101}{{\ttfamily arXiv:hep-th/9804101
  [hep-th]}}.

\bibitem{Pelaggi:2017wzr}
G.~M. Pelaggi, F.~Sannino, A.~Strumia, and E.~Vigiani, ``{Naturalness of
  asymptotically safe Higgs},''
  \href{http://dx.doi.org/10.3389/fphy.2017.00049}{{\em Front.in Phys.}
  {\bfseries 5} (2017) 49},
\href{http://arxiv.org/abs/1701.01453}{{\ttfamily arXiv:1701.01453 [hep-ph]}}.

\bibitem{Litim:2015iea}
D.~F. Litim, M.~Mojaza, and F.~Sannino, ``{Vacuum stability of asymptotically
  safe gauge-Yukawa theories},''
  \href{http://dx.doi.org/10.1007/JHEP01(2016)081}{{\em JHEP} {\bfseries 01}
  (2016) 081},
\href{http://arxiv.org/abs/1501.03061}{{\ttfamily arXiv:1501.03061 [hep-th]}}.

\bibitem{Caracciolo:1989pt}
S.~Caracciolo, G.~Curci, P.~Menotti, and A.~Pelissetto, ``{The Energy Momentum
  Tensor for Lattice Gauge Theories},''
\href{http://dx.doi.org/10.1016/0003-4916(90)90203-Z}{{\em Annals Phys.}
  {\bfseries 197} (1990) 119}.

\bibitem{Suzuki:2013gza}
H.~Suzuki, ``{Energy–momentum tensor from the Yang–Mills gradient flow},''
  \href{http://dx.doi.org/10.1093/ptep/ptt059, 10.1093/ptep/ptv094}{{\em PTEP}
  {\bfseries 2013} (2013) 083B03},
  \href{http://arxiv.org/abs/1304.0533}{{\ttfamily arXiv:1304.0533 [hep-lat]}}.
[Erratum: PTEP2015,079201(2015)].

\bibitem{DelDebbio:2013zaa}
L.~Del~Debbio, A.~Patella, and A.~Rago, ``{Space-time symmetries and the
  Yang-Mills gradient flow},''
  \href{http://dx.doi.org/10.1007/JHEP11(2013)212}{{\em JHEP} {\bfseries 11}
  (2013) 212},
\href{http://arxiv.org/abs/1306.1173}{{\ttfamily arXiv:1306.1173 [hep-th]}}.

\bibitem{Catterall:2009it}
S.~Catterall, D.~B. Kaplan, and M.~Unsal, ``{Exact lattice supersymmetry},''
  \href{http://dx.doi.org/10.1016/j.physrep.2009.09.001}{{\em Phys. Rept.}
  {\bfseries 484} (2009) 71--130},
\href{http://arxiv.org/abs/0903.4881}{{\ttfamily arXiv:0903.4881 [hep-lat]}}.

\bibitem{AFGJ97}
D.~Anselmi, D.~Freedman, M.~T. Grisaru, and A.~Johansen, ``{Nonperturbative
  formulas for central functions of supersymmetric gauge theories},''
  \href{http://dx.doi.org/10.1016/S0550-3213(98)00278-8}{{\em Nucl.Phys.}
  {\bfseries B526} (1998) 543--571},
\href{http://arxiv.org/abs/hep-th/9708042}{{\ttfamily arXiv:hep-th/9708042
  [hep-th]}}.

\bibitem{DelDebbio:2010jy}
L.~Del~Debbio and R.~Zwicky, ``{Scaling relations for the entire spectrum in
  mass-deformed conformal gauge theories},''
  \href{http://dx.doi.org/10.1016/j.physletb.2011.04.059}{{\em Phys. Lett.}
  {\bfseries B700} (2011) 217--220},
\href{http://arxiv.org/abs/1009.2894}{{\ttfamily arXiv:1009.2894 [hep-ph]}}.

\bibitem{Shore:2016xor}
G.~M. Shore, \href{http://dx.doi.org/10.1007/978-3-319-54000-9}{{\em {The c and
  a-theorems and the Local Renormalisation Group}}}.
\newblock Springerbriefs in physics. Springer, Cham, 2017.
\newblock
\href{http://arxiv.org/abs/1601.06662}{{\ttfamily arXiv:1601.06662 [hep-th]}}.
\newblock

\bibitem{H81}
S.~Hathrell, ``{Trace Anomalies and {QED} in Curved Space},''
\href{http://dx.doi.org/10.1016/0003-4916(82)90227-5}{{\em Annals Phys.}
  {\bfseries 142} (1982) 34}.

\bibitem{Kadanoff:1978pv}
L.~P. Kadanoff and A.~C. Brown, ``{Correlation functions on the critical lines
  of the Baxter and Ashkin-Teller models},''
\href{http://dx.doi.org/10.1016/0003-4916(79)90100-3}{{\em Annals Phys.}
  {\bfseries 121} (1979) 318--342}.

\end{thebibliography}\endgroup

\end{document}